\documentclass[aip,reprint,floatfix]{revtex4-1}

\newcommand{\be}{\begin{equation}}
\newcommand{\ee}{\end{equation}}
\newcommand{\bea}{\begin{align}}
\newcommand{\eea}{\end{align}}

\renewcommand{\Re}{\mathrm{Re}\,}

\newcommand{\Cu}{\rm Cu}
\newcommand{\Ti}{\rm Ti}

\newcommand{\mK}{\mathrm{mK}}

\usepackage{hyperref} 
\usepackage{graphicx}
\usepackage{amsmath, amsthm, amssymb}
\usepackage{epstopdf}
\usepackage{epsfig}
\usepackage{epsf}
\usepackage{array}
\usepackage{mathtools}
\usepackage{dcolumn}
\usepackage[tight]{subfigure}
\usepackage[normalem]{ulem}
\usepackage{verbatim}
\usepackage{inputenc}
\usepackage{color}
\usepackage{bm} 
\usepackage{natbib}
\usepackage{xspace}
\usepackage{mathbbol}
\usepackage{cleveref}
\DeclarePairedDelimiter\abs{\lvert}{\rvert}%

\begin{document}
\title{Negative differential thermal conductance by photonic transport in electronic circuits}

\author{Shobhit Saheb Dey}
\affiliation{Department of Physics, Indian Institute of Technology Kharagpur, Kharagpur, India 721302}
\affiliation{Quantum Research Centre, Technology Innovation Institute, Abu Dhabi, UAE}

\author{Giuliano Timossi}
\affiliation{NEST Istituto Nanoscienze-CNR and Scuola Normale Superiore, I-56127 Pisa, Italy}

\author{Luigi~Amico}
\affiliation{Quantum Research Centre, Technology Innovation Institute, Abu Dhabi, UAE}
\affiliation{Centre for Quantum Technologies, National University of Singapore, 3 Science Drive 2, Singapore 117543}
\affiliation{INFN-Sezione di Catania, Via S. Sofia 64, 95127 Catania, Italy}
\affiliation{LANEF `Chaire d’excellence’, Universit\'{e} Grenoble-Alpes \& CNRS, F-38000 Grenoble, France}
\affiliation{MajuLab, CNRS-UNS-NUS-NTU International Joint Research Unit, UMI 3654, Singapore}

\author{Giampiero Marchegiani}
\email{giampiero.marchegiani@tii.ae}
\affiliation{Quantum Research Centre, Technology Innovation Institute, Abu Dhabi, UAE}

\date{\today}

\setlength{\arraycolsep}{2pt}

\begin{abstract}
The negative differential thermal conductance (NDTC) provides  the key mechanism for  realizing  thermal transistors. This exotic effect has been the object of an extensive theoretical investigation, but the implementation is still limited to a few specific physical systems. Here, we consider a  simple circuit of  two electrodes exchanging heat through electromagnetic radiation. We demonstrate that the existence of an optimal condition for power transmission, well-known as impedance matching in electronics, provides a natural framework for engineering NDTC: the heat flux is reduced when the temperature increase is associated to an abrupt change of the electrode's impedance. As a case study, we analyze a hybrid structure based on thin-film technology, in which the increased resistance  is due to a superconductor-resistive phase transition. For typical metallic superconductors operating below $1$K, NDTC reflects  in a temperature drop of the order of a few mK by increasing the power supplied to the system. Our work draws new routes for implementing a thermal transistor in nanoscale circuits.
\end{abstract}

\date{\today}


\maketitle

The development of quantum technologies is  one of the main driving forces of the current physical research~\cite{Dowling2003,Acn2018}.
In many implementations of this discipline,  device miniaturization and low noise requirement have motivated an intense theoretical investigation and experimental activity on thermal transport in nanoscale solid-state devices~\cite{GiazottoRMP,DiVentraRMP2011,Bauer2012,Muhonen2012,CahillAPR2014,SongAIP2015,FornieriReview}. In this direction, as a combined effect of non equilibrium thermal fluctuations and non-linear response, physical systems in which  the heat flow $\dot Q$ decreases by increasing temperature gradients, i.e,  $d\dot Q/dT<0$  can be realized. In these regimes, the system is characterized by a  Negative  Differential Thermal Conductance (NDTC)~\cite{LiWangCasati2006}. Exploring the physical meaning of the NDTC regime defines a certainly interesting line of basic research~\cite{PhysRevB.76.020301,PhysRevE.81.041131,PhysRevB.80.104302,PhysRevE.87.062104,PhysRevE.81.041131}. At the same time, such effect allows to configure and fabricate new devices based on NDTC. An important case study in this context is provided by the thermal transistor introduced fifteen years ago by Casati and coworkers~\cite{LiWangCasati2006}. Over the years, several proposals for the implementation of thermal transistors and similar devices have been put forward, with technologies ranging from phononics~\cite{LiRMP84}, superconducting junctions~\cite{FornieriPRB93,Zare_2019}, electrochemical cells ~\cite{SoodNatComm2018} to near-field devices~\cite{Ben-Abdallah_PRL112,NearFieldRMP93,Moncada-Villa_PRApplied15}. Even  though great efforts have been devoted to the problem, engineering of NDTC in physical systems is still a challenging task, with   only  few experimental observations in specific systems  being carried out ~\cite{PhysRevX.3.041004,ItoAPL105,NbNPRB99}.
\begin{figure}[h]
    \centering
    \includegraphics[width=0.48\textwidth]{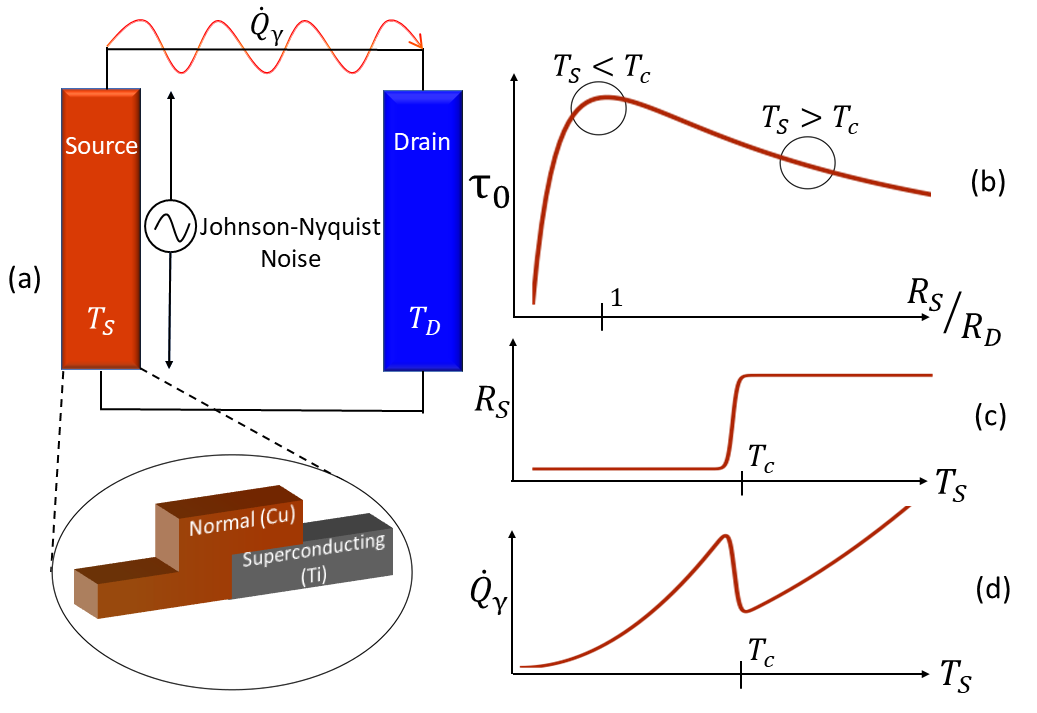}
    \caption{NDTC in photonic transport. (a) Circuit scheme: source and drain electrode are connected through a lossless line. In the presence of a thermal gradient, heat is transferred through a photon-mediated mechanism. In the blow-up, we show the implementation discussed in the second part of the manuscript, where the source is given as a series of a normal metal and a superconductor. (b) Power transfer coupling coefficient between two normal resistors $R_S$ and $R_D$. The transmission is maximum for $R_S=R_D$ (impedance matching condition).  (c) Sharp temperature increase of source resistance due to a phase-change. (d) Resulting NDTC in the photon-mediated heat current.}
    \label{Fig1}
\end{figure}
In our work, we will show  how a circuit approach to photon-mediated thermal transport can provide a general route for NDTC engineering. In our logic, we rely on  a specific 'resonant' property holding for the thermal transmission: 
The  electromagnetic power turns out to be  optimally transmitted between two circuital elements if a certain impedance matching condition is fulfilled~\cite{kikkert2013rf}. Starting from this condition, we impart an abrupt impedance mismatch   by a temperature change. We shall see that such protocol can lead to  NDTC. After a general discussion of the mechanism, we present a quantitative investigation in a realistic platform in which the impedance mismatch is achieved through the normal-to-superconductor phase transition.
We note that the photonic heat transport between two normal metals has been recently investigated in different experiments~\cite{MeschkePhotonic,RonzaniPhotonic,SeniorPhotonic,MailletPhotonic,Thomas_PhotonicJJ_PRB100,gubaydullin2021photonic}, where superconducting elements are exploited to realize a tunable coupling, both in the classical and in the quantum regime~\cite{Karimi_2017,Thomas_PhotonicJJ_PRB100}. Here, we investigate photon-mediated heat transport in superconductors that, because of their impedance's strong temperature dependence,  provide important examples of coherent networks  in which non-linear transport effects can take place~\cite{Nefzaoui2014,BosisioPRB93,Ordonez-Miranda2017,MarchegianiAPLPhotonic,Moncada-Villa_PRApplied15}.

We consider the scheme depicted in Fig.~\ref{Fig1}(a). The system is composed of two electrodes, denoted as source (S) and drain (D), electrically connected through wires of negligible losses. For the sake of generality, we follow a lumped parameters approach valid at low temperatures for  photons wavelengths larger than the size of the typical circuit element~\footnote{At $T=1$K, the photon thermal wavelength is $\lambda=hc/(k_B T)=1.4$ cm.}. We assume that the lead S is in thermal equilibrium with temperature $T_S$, while the drain resides at temperature $T_D$. When a thermal gradient is present, i.e., for $T_S\neq T_D$, heat is exchanged between the two elements. More precisely, the heat current generated due to electromagnetic fluctuations can be accounted for by means of the fluctuation-dissipation theorem~\cite{lifshitz2013statistical}. The Johnson-Nyquist voltage noise density in the source reads~\cite{SchmidtPRL93} $S_V(\omega)=4\omega\Re[Z_S(\omega,T_S)][n(\omega,T_S)+1/2] $, where $\omega$ is the photon energy, $n(\omega, T)=[e^{\omega/(k_BT)}-1]^{-1}$ is the Bose distribution, and
$Z_S(\omega,T_S)$ is the source's impedance. The corresponding noise current in the circuit is $S_I(\omega)=S_V(\omega)/|Z_{\rm tot}|^2$, where $Z_{\rm tot}=Z_S(\omega,T_S)+Z_D(\omega,T_D)$ is the total impedance of the circuit, and hence the power density dissipated in the drain is $S_P(\omega)=\Re[Z_D(\omega,T_D)] S_I(\omega)$. The total power transferred is obtained by integrating over the photon frequency $\nu=\omega/h$ ($h$ is the Planck's constant), giving
\begin{equation}
     \dot Q_{S\to D}(T_S,T_D) =\frac{1}{h} 
     \int_{0}^{\infty} \omega\tau(\omega,T_S,T_D) \left[n(\omega,T_S)+\frac{1}{2}\right] \,d\omega,
     \label{PowStoD}
\end{equation}
where the effective photon transmission coefficient has been identified as~\cite{SchmidtPRL93,OjanenPRL100,PascalPRB83}:
\begin{equation}
    \tau(\omega,T_S,T_D) = 4\frac{\text{Re[}Z_S(\omega,T_S)]\text{Re[}Z_D(\omega,T_D)]}{\abs{Z_S(\omega,T_S) + Z_D(\omega,T_D)}^2}.
    \label{tau}
\end{equation}
The net power transmitted is then obtained by subtracting the heat current $\dot Q_{D\to S}$ flowing from the drain to the source, i.e., $\dot Q_\gamma=\dot Q_{S\to D}-\dot Q_{D\to S}$. Due to symmetry, $\dot Q_{D\to S}$ is simply obtained by exchanging $S\leftrightarrow D$ in Eq.~\eqref{PowStoD}, yielding the following expression for $\dot Q_\gamma$~\cite{MeschkePhotonic,PascalPRB83}:
\begin{equation}
     \dot Q_{\gamma}(T_S,T_D) =\frac{1}{h} \int_{0}^{\infty} \omega\tau(\omega,T_S,T_D) [n(\omega,T_S)-n(\omega,T_D)] \,d\omega.
     \label{Landauers}
\end{equation}
In this setting, a NDTC can be achieved, for instance, when $\dot{Q}_\gamma$~\textit{decreases} by increasing $T_S$ for a fixed value of $T_D<T_S$. To give a simplified view, we first discuss the case where $Z_S,Z_D$ have no reactive components and do not depend on the photon energy, i.e., $Z_S(\omega,T_S)=R_S(T_S)$ and $Z_D(\omega,T_D)=R_D(T_D)$. In this case, the integral in Eq.~\eqref{Landauers} can be explicitly evaluated as $\dot{Q}_\gamma(T_S,T_D) = \tau_0(R_S/R_D)\pi^2k_B^2(T_S^2 - T_D^2)/(6 h)$, where $\tau_0(x)=4x/(x+1)^2$.
 
The key property that we use to engineer the NDTC is  the non monotonous behaviour of the transmission $\tau_0$  as function of the  resistance ratio $(R_S/R_D)$; in particular  $\tau_0$ results to be  maximum at the impedance matching condition, i.e., for $R_S=R_D$~\cite{PascalPRB83}. This result is also known in the literature as the maximum power transfer theorem: it expresses a  general statement on optimal power transfer between two elements, ranging from mechanical collisions to electromagnetic phenomena (such as the one investigated here)~\cite{Harrison2013,Atkin2013}. Therefore, an abrupt change of $R_S(T_S)$ at the critical temperature $T_c$, idealized as a resistance jump as in Fig.~\ref{Fig1} (c), makes $\dot Q_{\gamma}$ to decrease by increasing $T_S$ -  Fig.~\ref{Fig1} (d). In a more realistic setting, the resistance jump results to be  smoothed out in a finite temperature range. 

Now we discuss a possible scheme for the detection and implementation of NDTC. To mimic the sharp resistance increase of Fig.~\ref{Fig1}c, we exploit a superconducting to normal phase transition. The drain electrode is a normal resistor with resistance $R_D$. We assume the source is composed of a series of a normal element (with resistance $R_N=R_D$ to ensure impedance matching) and a superconducting element, connected through a clean contact of negligible resistance (see Fig.~\ref{Fig1}a)~\footnote{For simplicity, we neglect the superconducting proximity effect between the two elements~\cite{Pannetier2000}.}. Below the superconducting critical temperature, the dissipation in the superconductor is exponentially suppressed, giving $\Re[Z_S(\omega,T_S)]\sim R_D$, whereas $\Re[Z_S(\omega,T_S)]= R_D+R_0$ for $T>T_c$, where $R_0$ is the resistance of the superconducting element in the normal state. We stress that this mechanism is generally related to the impedance mismatch generated by the superconducting to normal transition, and holds well beyond the more specific near-field regime experimentally reported in Ref.~\onlinecite{NbNPRB99}. We choose to operate in the sub-Kelvin regime, where the photonic heat current is more relevant to the thermal equilibration~\cite{SchmidtPRL93,BosisioPRB93,MarchegianiAPLPhotonic}. On the material side, we consider thin metallic films which can be deposited through electron-beam evaporation, such as Titanium (Ti) as the superconductor (with typical critical temperature in the range $0.3-0.5$K~\cite{TitaniumPhysRev92,DresselTitanium}) and copper (Cu) for the normal conducting elements. The wires can be realized with superconducting aluminum (Al), which displays higher critical temperature than Ti~\footnote{The bulk critical temperature of Al is $T_c^{\rm Al}=1.2$ K and larger for thin films (up to $3K$ for 3 nm thick films).},  and therefore can account for realizing  lossless lines~\footnote{The dissipation in the wire is exponentially suppressed at low temperatures due to the presence of the gap. The kinetic inductance contribution can be minimized by geometry design. At the same time, direct conduction of heat between the electrodes and the wire is suppressed by Andreev mirroring.~\cite{MarchegianiAPLPhotonic}} (as required in our model).

We start by discussing the photon-mediated heat transport.  In our setup, the impedance of the two leads are $Z_S(\omega,T_S)=R_D+Z_0(\omega,T_S)$, and $Z_D(\omega,T_S)=R_D$. As recently experimentally demonstrated~\cite{DresselTitanium}, the complex impedance of Ti can be modeled within the Mattis-Bardeen theory~\cite{MattisBardeenPhysRev111}, valid for Bardeen-Cooper-Schrieffer (BCS) superconductors. More precisely, the complex impedance reads $Z_0(\omega,T_S)=R_{0}[\sigma_1(\omega,T_S) - \iota \sigma_2(\omega,T_S)]^{-1}$, where the real ($\sigma_1$) and the imaginary ($\sigma_2$) parts of the complex conductivity (scaled to the normal conductivity) are expressed by:
\begin{align}
    \sigma_1(\omega,T)=\frac{2}{\omega}\int_{\Delta}^{\infty}\nu(E,E')[f(E,T) - f(E',T)] \, dE \nonumber\\ -
    \frac{\Theta(\omega - 2\Delta)}{\omega}\int_{\Delta - \omega}^{-\Delta}\nu(E,E')[1 - 2f(E',T)]\, dE, \label{eq:MBreal}\\
    \sigma_2(\omega,T)= \frac{1}{\omega}\int_{-\Delta, \Delta-\omega}^{\Delta}|\nu(E,E')|[1 - 2f(E',T)] \, dE.
    \label{eq:MBimag}
\end{align}
Above, $\Theta(x)$ is the Heaviside-step function, $E'=E+\omega$, $f(E,T)=[e^{E/(k_BT)}+1]^{-1}$ is the Fermi function, and we introduced the function~\footnote{In Eq.~\eqref{eq:MBimag}, $\sqrt{E^2-\Delta^2}$ is purely imaginary and should be intended as $\sqrt{\Delta^2-E^2}$ due to the presence of the absolute value. The lower limit of integration is the maximum between $-\Delta$ and $\Delta-\omega$.}
\begin{equation}
\nu(E,E')=\frac{E E'+\Delta^2}{\sqrt{E^2-\Delta^2}\sqrt{E'^2-\Delta^2}}.
\end{equation}
The superconducting gap is approximately given by $\Delta(T) = \Delta_0\tanh(1.74\sqrt{T_c/T -1})$~\footnote{This approximate expression gives an accurate description for the temperature dependence of the gap, with an error smaller than 3\% for every temperature.} where $\Delta_0= 1.76 k_BT_c$, according to the BCS theory. The dissipative processes in the superconductor are given by thermally excited quasiparticles (first integral of Eq.~\eqref{eq:MBreal}) or through pair-breaking processes due to  photon absorption that may occur for   $\omega>2\Delta$ (second integral of Eq.~\eqref{eq:MBreal}). The imaginary part of the conductivity Eq.~\eqref{eq:MBimag} gives the kinetic inductance of the superconducting film. 

\begin{figure}[h]
    \centering
    \includegraphics[width = \linewidth]{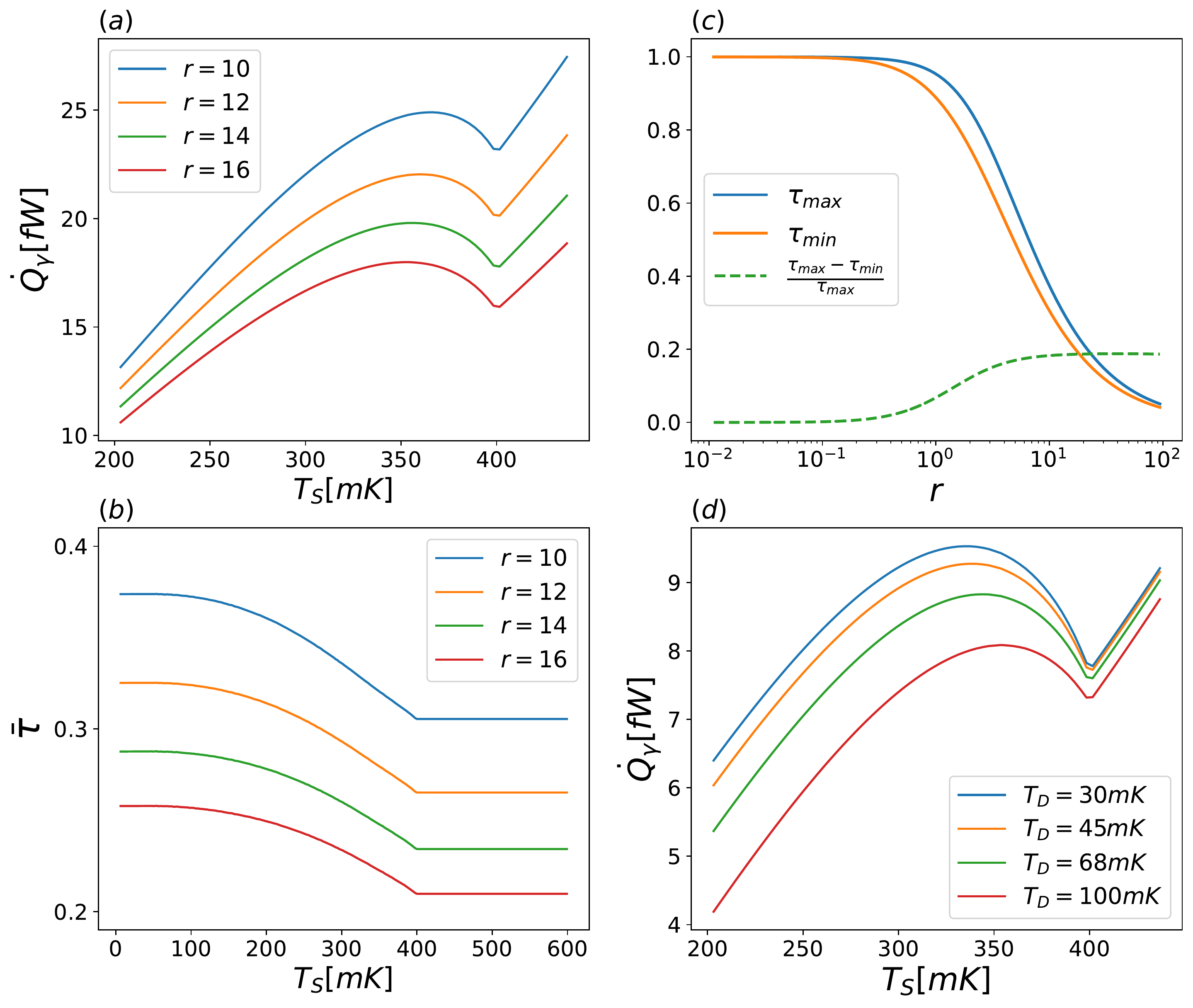}
    \caption{NDTC in the specific implementation based on low-temperature superconductors. (a)  $\dot{Q}_\gamma$ vs $T_S$ with $T_D=30$mK for different values of $r=R_0/R_D$. NDTC appears for source temperatures $T_S$ around $(350{\rm mK,400mK})$, (b) Energy averaged thermal transmission $\bar\tau$ vs $T_S$ for the same values of $r$ as in (a), (c) $\tau_{max}$, $\tau_{min}$ and their relative difference as function of $r$ showing favourable operating region to observe NDTC, (d) $\dot{Q}_\gamma$ vs $T_S$ for $r=36$ and different values of $T_D$.
    \label{Fig2}
    }
\end{figure}

Figure~\ref{Fig2}a displays $\dot Q_\gamma$ as a function of the source temperature $T_S$, for $T_D=30$mK and different values of the ratio $r=R_0/R_D$. Notably, the evolution is non-monotonic with $T_S$, and the system displays NDTC for source temperatures approximately in the range around $[365\mK,400\mK]$. Upon increasing $R_0$, the photonic heat current decreases. Here the reduction of $\dot Q_\gamma$ with $R_0$ for $T>T_c$ arises from  suppression of  $\tau_0(x)$ for $x>1$, while  for $T<T_c$ it is related to the imaginary part of the $Z_S(\omega,T_S)$ i.e kinetic inductance. To analyze  such a  feature, we focus on the temperature dependence  of $\tau$. We define an average transmission $\bar\tau(T_S)=(2\Delta_0)^{-1}\int_{0}^{2\Delta_0} \tau(\omega,T_S)d\omega$~\footnote{The cutoff at $2\Delta_0$ is related to the fact that we are mainly interested in temperatures $T_S\leq T_c$.}. 
Figure~\ref{Fig2}b displays $\bar\tau$ as a function of the source temperature $T_S$, for the same values of $r$ as in Fig.~\ref{Fig2}a. The transmission is monotonically decreasing with $T_S$, being maximum at $T_S\ll T_c$, i.e., $\tau_{\rm max}=\bar\tau(T_S\to 0)$, and constant for $T_S\geq T_c$, with $\tau_{\rm min}=4 (1+r)/(2+r)^2$. Differently from the idealized situation displayed in Fig.~\ref{Fig1}, $\tau<1$ even at low temperatures. This behaviour is due to the non-negligible kinetic inductance of the superconductor, which reduces $\tau$  (see Eq.~\eqref{tau}). Hence, even though an increase of $R_0$ turns out always beneficial in the schematics of Fig.~\ref{Fig1}b, in the realistic case, an arbitrarily large value of $R_0$ may hinder the NDTC behavior as also seen in Fig.~\ref{Fig2}a. Indeed, $\tau_{\rm max}$ and $\tau_{\rm min}$ decreases monotonically as a function of $r$, as displayed in Fig.~\ref{Fig2}c (solid curves). Notably, the relative transmission modulation, defined as $(\tau_{\rm max}-\tau_{\rm min})/\tau_{\rm max}$, grows monotonically with $r$, and it is approximately saturating to $\sim0.2$ for $r\sim 10$ (see dashed curve in Fig.~\ref{Fig2}c). As a result, the phenomenology of NDTC is well displayed for values of $r$ in the range [10,100].

In Fig.~\ref{Fig2}d, we display $\dot Q_\gamma$ as a function of $T_S$ for $r=36$ and different values of $T_D$. For a given value of the source temperature, the heat current typically decreases by increasing $T_D$, due to the reduction of the thermal gradient. Notably, the NDTC phenomenology for source temperatures $T_S\lesssim T_c$ is robust even with a sizeable change of the drain temperature. This feature is crucial for the detection of the NDTC discussed below.

In the above computations, we characterized the photon-mediated thermal transport as a function of the temperatures of the source for a selected values of the drain temperature. However, in an actual experiment, the temperatures $T_S$ and $T_D$ are not directly controlled. Instead, power is injected into the system and the temperature $T_S,T_D$ are the result of the power balance in each electrode~\cite{FornieriReview}. More precisely, experiments based on low-temperature superconducting thin films are typically well described within the quasi-equilibrium regime~\cite{GiazottoRMP,Muhonen2012}. Namely, in each electrode, quasiparticles and phonons can be treated as separate subsystems, which may thermalize to different temperatures, since the electron-phonon scattering rate slows down at very low temperatures.  

\begin{figure}[h]
    \centering
    \includegraphics[width = \linewidth]{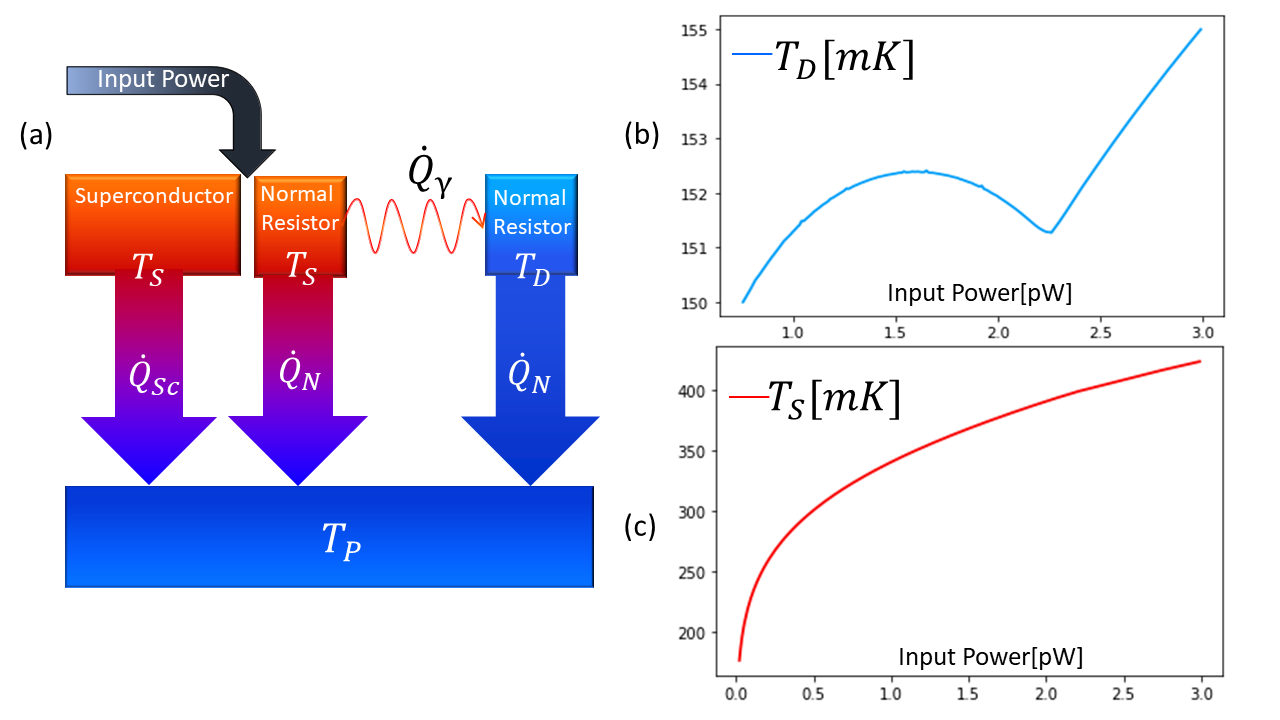}
    \caption{Thermal balance and source and drain steady-state temperatures. (a) Heat flow diagram: input power is provided to the source, and losses due to electron-phonon coupling are included. (b) Steady state temperature of the drain $T_D$ vs input power $P_{\rm in}$, showing negative differential characteristics with respect to $P_{\rm in}$. (c) Steady state temperature of the source $T_S$ vs $P_{\rm in}$, showing a monotonic increase with input power. Parameters: $T_p=30$mK, $\rho_{\Cu}=3\mu\Omega\cdot $cm~\cite{ViisanenPRB97}, $l_{\Cu}=2\ \mu$m, $\mathcal A_{\Cu} =700\times 30$ nm$^2$, $\rho_{\Ti}=30\mu\Omega\cdot $cm~\cite{DeSimoniGate}, $l_{\Ti}=6\mu$m, $w_{\Ti}=700\times 25$nm$^2$,  where $\rho,l,\mathcal A$ are the normal state resistivity, the length and the cross section of the films, respectively. 
    \label{Fig3}
   }
\end{figure}

The thermal exchanges in our system are schematically depicted in Fig.~\ref{Fig3}a. We assume that the phonons in each part of the device are well thermalized to the substrate temperature ($T_p\sim 30$ mK  set by the cryogenic dilution fridge). Due to the clean contact, we neglect any potential small thermal gradient in the source electrode, characterizing the electron (in the normal metal) and the quasiparticle (in the superconductor) subsystem with a single temperature $T_S$. The electronic temperature of the drain is given by $T_D$. When power $P_{\rm in}$ is injected in the source, the source temperature increases $T_S>T_p$, producing photon-mediated exchange $\dot Q_\gamma$ with the drain. At the same time, heat is exchanged with the phononic bath both in the normal ($\dot Q_N$) and in the superconducting ($\dot Q_{Sc}$) parts of the device. At the steady state condition, in each element the ingoing heat current must be equal to the outgoing heat current, giving:
\begin{equation}
\begin{cases}
P_{\rm in}=\dot{Q}_\gamma(T_S,T_D) + \dot{Q}_N(T_S,T_p) + \dot{Q}_{Sc}(T_S,T_p) \\
\dot{Q}_\gamma(T_S,T_D) = \dot{Q}_N(T_D,T_p)
\end{cases}
\label{equilibrium}
\end{equation}
The electron-phonon coupling in each Cu lead is given by $\dot Q_{N}(T,T_p)=\Sigma_{\rm Cu} \mathcal V_N(T^5-T_p^5)$, where $\mathcal V_N$ is the volume of each copper electrode and $\Sigma_{\rm Cu}=3\times10^9  {\rm W\ m}^{-3} {\rm K}^{-5}$ is the material dependent electron-phonon coupling constant~\cite{GiazottoRMP}. For the superconductor, the electron-phonon interaction reads~\cite{TimofeevPRL102,MaisiPRL111}
\begin{align}
    \dot{Q}_{Sc}(T_S, T_{p}) = \lambda_s \int_{0}^{\infty} d\omega\omega^3 [n(\omega,T_S) - n(\omega,T_{p})] \mathcal F(\omega,T_S), \label{eph_general} 
\end{align}
where $\lambda_s=\Sigma_{\rm Ti} \mathcal V/(24 \zeta(5) k_B^5)$ and
\begin{align}
    \mathcal F(\omega,T_S)=\int_{-\infty}^{\infty}dE\rho(E)\rho(E')\left(1-\frac{\Delta^2}{E E'}\right)\nonumber\\ [f(E,T_S) - f(E',T_S)]
\end{align}
with $E'=E+\omega$.
Above, $\Sigma_{\Ti}=1.3\times10^9  {\rm W\ m}^{-3} {\rm K}^{-5}$ is the coupling coefficient~\cite{GiazottoRMP}, $\mathcal V$ is the volume of the superconducting lead, $\zeta(z)$ is the Riemann-Zeta function, and $\rho(E)=|E|\theta(|E|-\Delta)/\sqrt{E^2-\Delta^2}$ is the BCS density of states. 
For given values of the phonon temperature $T_p$ and the input power $P_{\rm in}$  Eq.~\eqref{equilibrium} is a nonlinear system of integro-algebraic equations in the two-variables for $T_S$ and $T_D$. 

Figure~\ref{Fig3}(b)-(c) displays the solution of Eq.~\eqref{equilibrium}, obtained through a Newton-Raphson optimization algorithm, as a function of the input power $P_{\rm in}$. Notably, while $T_S$ increases monotonously with $P_{\rm in}$ (see Fig.~\ref{Fig3}(c)), the drain temperature $T_D$ decreases for input power around the range $[1.5,2.25]$ pW (see Fig.~\ref{Fig3}(b)). This behavior is a signature of NDTC in the photonic channel. Indeed, when $T_S$ is close to $T_c$ ($~400$ mK) the heat exchange is reduced by increasing $P_{\rm in}$, in agreement with our previous discussion, resulting in a decrease of $T_D$ due to the second equation in Eq.~\eqref{equilibrium}.

In summary, we investigated the Negative Differential Thermal Conductance (NDTC) in the photonic heat transport between two electrodes, within a lumped circuital approach. We rely on a photon resonant transmission that, in analogy of the electronic principle of impedance matching~\cite{kikkert2013rf}, minimize the loss in the heat transmission.  We demonstrated how such resonant transmission can be exploited as a reference point to realize NDTC. An abrupt impedance mismatch causes a reduction of the heat flow that can be transmitted among the two electrodes, and therefore the differential conductance is negative. In the second part of the paper, we focus on a specific design based on low-temperature superconductor, that can be realized with the state-of-the-art nano-fabrication techniques. Our calculations shows that the effect can be  identified with well-established thermometric techniques, leading to temperature drops larger than $1$mK. Our work provides a general protocol for engineering NDTC in electric circuits where heat is exchanged through electromagnetic radiation. In this respect, superconducting circuits certainly provide a promising platform. The general nature of the maximum power transfer condition, though, may trigger investigations of NDTC on a wider variety of physical implementations.     

We thank Gianluigi Catelani, and Alessandro Braggio for fruitful discussions.

\section*{Data Availability}
The data that support the findings of this study are available from the corresponding author upon reasonable request.


\begin{thebibliography}{57}%
\makeatletter
\providecommand \@ifxundefined [1]{%
 \@ifx{#1\undefined}
}%
\providecommand \@ifnum [1]{%
 \ifnum #1\expandafter \@firstoftwo
 \else \expandafter \@secondoftwo
 \fi
}%
\providecommand \@ifx [1]{%
 \ifx #1\expandafter \@firstoftwo
 \else \expandafter \@secondoftwo
 \fi
}%
\providecommand \natexlab [1]{#1}%
\providecommand \enquote  [1]{``#1''}%
\providecommand \bibnamefont  [1]{#1}%
\providecommand \bibfnamefont [1]{#1}%
\providecommand \citenamefont [1]{#1}%
\providecommand \href@noop [0]{\@secondoftwo}%
\providecommand \href [0]{\begingroup \@sanitize@url \@href}%
\providecommand \@href[1]{\@@startlink{#1}\@@href}%
\providecommand \@@href[1]{\endgroup#1\@@endlink}%
\providecommand \@sanitize@url [0]{\catcode `\\12\catcode `\$12\catcode
  `\&12\catcode `\#12\catcode `\^12\catcode `\_12\catcode `\%12\relax}%
\providecommand \@@startlink[1]{}%
\providecommand \@@endlink[0]{}%
\providecommand \url  [0]{\begingroup\@sanitize@url \@url }%
\providecommand \@url [1]{\endgroup\@href {#1}{\urlprefix }}%
\providecommand \urlprefix  [0]{URL }%
\providecommand \Eprint [0]{\href }%
\providecommand \doibase [0]{http://dx.doi.org/}%
\providecommand \selectlanguage [0]{\@gobble}%
\providecommand \bibinfo  [0]{\@secondoftwo}%
\providecommand \bibfield  [0]{\@secondoftwo}%
\providecommand \translation [1]{[#1]}%
\providecommand \BibitemOpen [0]{}%
\providecommand \bibitemStop [0]{}%
\providecommand \bibitemNoStop [0]{.\EOS\space}%
\providecommand \EOS [0]{\spacefactor3000\relax}%
\providecommand \BibitemShut  [1]{\csname bibitem#1\endcsname}%
\let\auto@bib@innerbib\@empty
\bibitem [{\citenamefont {Dowling}\ and\ \citenamefont
  {Milburn}(2003)}]{Dowling2003}%
  \BibitemOpen
  \bibfield  {author} {\bibinfo {author} {\bibfnamefont {J.~P.}\ \bibnamefont
  {Dowling}}\ and\ \bibinfo {author} {\bibfnamefont {G.~J.}\ \bibnamefont
  {Milburn}},\ }\bibfield  {title} {\enquote {\bibinfo {title} {Quantum
  technology: the second quantum revolution},}\ }\href {\doibase
  10.1098/rsta.2003.1227} {\bibfield  {journal} {\bibinfo  {journal} {Phil.
  Trans. R. Soc. A.}\ }\textbf {\bibinfo {volume} {361}},\ \bibinfo {pages}
  {1655} (\bibinfo {year} {2003})}\BibitemShut {NoStop}%
\bibitem [{\citenamefont {Ac{\'{\i}}n}\ \emph {et~al.}(2018)\citenamefont
  {Ac{\'{\i}}n}, \citenamefont {Bloch}, \citenamefont {Buhrman}, \citenamefont
  {Calarco}, \citenamefont {Eichler}, \citenamefont {Eisert}, \citenamefont
  {Esteve}, \citenamefont {Gisin}, \citenamefont {Glaser}, \citenamefont
  {Jelezko} \emph {et~al.}}]{Acn2018}%
  \BibitemOpen
  \bibfield  {author} {\bibinfo {author} {\bibfnamefont {A.}~\bibnamefont
  {Ac{\'{\i}}n}}, \bibinfo {author} {\bibfnamefont {I.}~\bibnamefont {Bloch}},
  \bibinfo {author} {\bibfnamefont {H.}~\bibnamefont {Buhrman}}, \bibinfo
  {author} {\bibfnamefont {T.}~\bibnamefont {Calarco}}, \bibinfo {author}
  {\bibfnamefont {C.}~\bibnamefont {Eichler}}, \bibinfo {author} {\bibfnamefont
  {J.}~\bibnamefont {Eisert}}, \bibinfo {author} {\bibfnamefont
  {D.}~\bibnamefont {Esteve}}, \bibinfo {author} {\bibfnamefont
  {N.}~\bibnamefont {Gisin}}, \bibinfo {author} {\bibfnamefont {S.~J.}\
  \bibnamefont {Glaser}}, \bibinfo {author} {\bibfnamefont {F.}~\bibnamefont
  {Jelezko}},  \emph {et~al.},\ }\bibfield  {title} {\enquote {\bibinfo {title}
  {The quantum technologies roadmap: a european community view},}\ }\href
  {\doibase 10.1088/1367-2630/aad1ea} {\bibfield  {journal} {\bibinfo
  {journal} {New J. Phys.}\ }\textbf {\bibinfo {volume} {20}},\ \bibinfo
  {pages} {080201} (\bibinfo {year} {2018})}\BibitemShut {NoStop}%
\bibitem [{\citenamefont {Giazotto}\ \emph {et~al.}(2006)\citenamefont
  {Giazotto}, \citenamefont {Heikkil\"a}, \citenamefont {Luukanen},
  \citenamefont {Savin},\ and\ \citenamefont {Pekola}}]{GiazottoRMP}%
  \BibitemOpen
  \bibfield  {author} {\bibinfo {author} {\bibfnamefont {F.}~\bibnamefont
  {Giazotto}}, \bibinfo {author} {\bibfnamefont {T.~T.}\ \bibnamefont
  {Heikkil\"a}}, \bibinfo {author} {\bibfnamefont {A.}~\bibnamefont
  {Luukanen}}, \bibinfo {author} {\bibfnamefont {A.~M.}\ \bibnamefont {Savin}},
  \ and\ \bibinfo {author} {\bibfnamefont {J.~P.}\ \bibnamefont {Pekola}},\
  }\bibfield  {title} {\enquote {\bibinfo {title} {Opportunities for
  mesoscopics in thermometry and refrigeration: Physics and applications},}\
  }\href {\doibase 10.1103/RevModPhys.78.217} {\bibfield  {journal} {\bibinfo
  {journal} {Rev. Mod. Phys.}\ }\textbf {\bibinfo {volume} {78}},\ \bibinfo
  {pages} {217--274} (\bibinfo {year} {2006})}\BibitemShut {NoStop}%
\bibitem [{\citenamefont {Dubi}\ and\ \citenamefont
  {Di~Ventra}(2011)}]{DiVentraRMP2011}%
  \BibitemOpen
  \bibfield  {author} {\bibinfo {author} {\bibfnamefont {Y.}~\bibnamefont
  {Dubi}}\ and\ \bibinfo {author} {\bibfnamefont {M.}~\bibnamefont
  {Di~Ventra}},\ }\bibfield  {title} {\enquote {\bibinfo {title} {Colloquium:
  Heat flow and thermoelectricity in atomic and molecular junctions},}\ }\href
  {\doibase 10.1103/RevModPhys.83.131} {\bibfield  {journal} {\bibinfo
  {journal} {Rev. Mod. Phys.}\ }\textbf {\bibinfo {volume} {83}},\ \bibinfo
  {pages} {131} (\bibinfo {year} {2011})}\BibitemShut {NoStop}%
\bibitem [{\citenamefont {Bauer}, \citenamefont {Saitoh},\ and\ \citenamefont
  {van Wees}(2012)}]{Bauer2012}%
  \BibitemOpen
  \bibfield  {author} {\bibinfo {author} {\bibfnamefont {G.~E.~W.}\
  \bibnamefont {Bauer}}, \bibinfo {author} {\bibfnamefont {E.}~\bibnamefont
  {Saitoh}}, \ and\ \bibinfo {author} {\bibfnamefont {B.~J.}\ \bibnamefont {van
  Wees}},\ }\bibfield  {title} {\enquote {\bibinfo {title} {Spin
  caloritronics},}\ }\href {\doibase 10.1038/nmat3301} {\bibfield  {journal}
  {\bibinfo  {journal} {Nat. Mater.}\ }\textbf {\bibinfo {volume} {11}},\
  \bibinfo {pages} {391--399} (\bibinfo {year} {2012})}\BibitemShut {NoStop}%
\bibitem [{\citenamefont {Muhonen}, \citenamefont {Meschke},\ and\
  \citenamefont {Pekola}(2012)}]{Muhonen2012}%
  \BibitemOpen
  \bibfield  {author} {\bibinfo {author} {\bibfnamefont {J.~T.}\ \bibnamefont
  {Muhonen}}, \bibinfo {author} {\bibfnamefont {M.}~\bibnamefont {Meschke}}, \
  and\ \bibinfo {author} {\bibfnamefont {J.~P.}\ \bibnamefont {Pekola}},\
  }\bibfield  {title} {\enquote {\bibinfo {title} {Micrometre-scale
  refrigerators},}\ }\href {\doibase 10.1088/0034-4885/75/4/046501} {\bibfield
  {journal} {\bibinfo  {journal} {Rep. Prog. Phys.}\ }\textbf {\bibinfo
  {volume} {75}},\ \bibinfo {pages} {046501} (\bibinfo {year}
  {2012})}\BibitemShut {NoStop}%
\bibitem [{\citenamefont {Cahill}\ \emph {et~al.}(2014)\citenamefont {Cahill},
  \citenamefont {Braun}, \citenamefont {Chen}, \citenamefont {Clarke},
  \citenamefont {Fan}, \citenamefont {Goodson}, \citenamefont {Keblinski},
  \citenamefont {King}, \citenamefont {Mahan}, \citenamefont {Majumdar},
  \citenamefont {Maris}, \citenamefont {Phillpot}, \citenamefont {Pop},\ and\
  \citenamefont {Shi}}]{CahillAPR2014}%
  \BibitemOpen
  \bibfield  {author} {\bibinfo {author} {\bibfnamefont {D.~G.}\ \bibnamefont
  {Cahill}}, \bibinfo {author} {\bibfnamefont {P.~V.}\ \bibnamefont {Braun}},
  \bibinfo {author} {\bibfnamefont {G.}~\bibnamefont {Chen}}, \bibinfo {author}
  {\bibfnamefont {D.~R.}\ \bibnamefont {Clarke}}, \bibinfo {author}
  {\bibfnamefont {S.}~\bibnamefont {Fan}}, \bibinfo {author} {\bibfnamefont
  {K.~E.}\ \bibnamefont {Goodson}}, \bibinfo {author} {\bibfnamefont
  {P.}~\bibnamefont {Keblinski}}, \bibinfo {author} {\bibfnamefont {W.~P.}\
  \bibnamefont {King}}, \bibinfo {author} {\bibfnamefont {G.~D.}\ \bibnamefont
  {Mahan}}, \bibinfo {author} {\bibfnamefont {A.}~\bibnamefont {Majumdar}},
  \bibinfo {author} {\bibfnamefont {H.~J.}\ \bibnamefont {Maris}}, \bibinfo
  {author} {\bibfnamefont {S.~R.}\ \bibnamefont {Phillpot}}, \bibinfo {author}
  {\bibfnamefont {E.}~\bibnamefont {Pop}}, \ and\ \bibinfo {author}
  {\bibfnamefont {L.}~\bibnamefont {Shi}},\ }\bibfield  {title} {\enquote
  {\bibinfo {title} {Nanoscale thermal transport. ii. 2003–2012},}\ }\href
  {\doibase 10.1063/1.4832615} {\bibfield  {journal} {\bibinfo  {journal}
  {Appl. Phys. Rev.}\ }\textbf {\bibinfo {volume} {1}},\ \bibinfo {pages}
  {011305} (\bibinfo {year} {2014})}\BibitemShut {NoStop}%
\bibitem [{\citenamefont {Song}\ \emph {et~al.}(2015)\citenamefont {Song},
  \citenamefont {Fiorino}, \citenamefont {Meyhofer},\ and\ \citenamefont
  {Reddy}}]{SongAIP2015}%
  \BibitemOpen
  \bibfield  {author} {\bibinfo {author} {\bibfnamefont {B.}~\bibnamefont
  {Song}}, \bibinfo {author} {\bibfnamefont {A.}~\bibnamefont {Fiorino}},
  \bibinfo {author} {\bibfnamefont {E.}~\bibnamefont {Meyhofer}}, \ and\
  \bibinfo {author} {\bibfnamefont {P.}~\bibnamefont {Reddy}},\ }\bibfield
  {title} {\enquote {\bibinfo {title} {Near-field radiative thermal transport:
  From theory to experiment},}\ }\href {\doibase 10.1063/1.4919048} {\bibfield
  {journal} {\bibinfo  {journal} {AIP Adv.}\ }\textbf {\bibinfo {volume} {5}},\
  \bibinfo {pages} {053503} (\bibinfo {year} {2015})}\BibitemShut {NoStop}%
\bibitem [{\citenamefont {Fornieri}\ and\ \citenamefont
  {Giazotto}(2017)}]{FornieriReview}%
  \BibitemOpen
  \bibfield  {author} {\bibinfo {author} {\bibfnamefont {A.}~\bibnamefont
  {Fornieri}}\ and\ \bibinfo {author} {\bibfnamefont {F.}~\bibnamefont
  {Giazotto}},\ }\bibfield  {title} {\enquote {\bibinfo {title} {Towards
  phase-coherent caloritronics in superconducting circuits},}\ }\href {\doibase
  10.1038/nnano.2017.204} {\bibfield  {journal} {\bibinfo  {journal} {Nat.
  Nanotechnol.}\ }\textbf {\bibinfo {volume} {12}},\ \bibinfo {pages} {944}
  (\bibinfo {year} {2017})}\BibitemShut {NoStop}%
\bibitem [{\citenamefont {Li}, \citenamefont {Wang},\ and\ \citenamefont
  {Casati}(2006)}]{LiWangCasati2006}%
  \BibitemOpen
  \bibfield  {author} {\bibinfo {author} {\bibfnamefont {B.}~\bibnamefont
  {Li}}, \bibinfo {author} {\bibfnamefont {L.}~\bibnamefont {Wang}}, \ and\
  \bibinfo {author} {\bibfnamefont {G.}~\bibnamefont {Casati}},\ }\bibfield
  {title} {\enquote {\bibinfo {title} {Negative differential thermal resistance
  and thermal transistor},}\ }\href {\doibase 10.1063/1.2191730} {\bibfield
  {journal} {\bibinfo  {journal} {Appl. Phys. Lett.}\ }\textbf {\bibinfo
  {volume} {88}},\ \bibinfo {pages} {143501} (\bibinfo {year}
  {2006})}\BibitemShut {NoStop}%
\bibitem [{\citenamefont {Yang}\ \emph {et~al.}(2007)\citenamefont {Yang},
  \citenamefont {Li}, \citenamefont {Wang},\ and\ \citenamefont
  {Li}}]{PhysRevB.76.020301}%
  \BibitemOpen
  \bibfield  {author} {\bibinfo {author} {\bibfnamefont {N.}~\bibnamefont
  {Yang}}, \bibinfo {author} {\bibfnamefont {N.}~\bibnamefont {Li}}, \bibinfo
  {author} {\bibfnamefont {L.}~\bibnamefont {Wang}}, \ and\ \bibinfo {author}
  {\bibfnamefont {B.}~\bibnamefont {Li}},\ }\bibfield  {title} {\enquote
  {\bibinfo {title} {Thermal rectification and negative differential thermal
  resistance in lattices with mass gradient},}\ }\href {\doibase
  10.1103/PhysRevB.76.020301} {\bibfield  {journal} {\bibinfo  {journal} {Phys.
  Rev. B}\ }\textbf {\bibinfo {volume} {76}},\ \bibinfo {pages} {020301}
  (\bibinfo {year} {2007})}\BibitemShut {NoStop}%
\bibitem [{\citenamefont {He}\ \emph {et~al.}(2010)\citenamefont {He},
  \citenamefont {Ai}, \citenamefont {Chan},\ and\ \citenamefont
  {Hu}}]{PhysRevE.81.041131}%
  \BibitemOpen
  \bibfield  {author} {\bibinfo {author} {\bibfnamefont {D.}~\bibnamefont
  {He}}, \bibinfo {author} {\bibfnamefont {B.-q.}\ \bibnamefont {Ai}}, \bibinfo
  {author} {\bibfnamefont {H.-K.}\ \bibnamefont {Chan}}, \ and\ \bibinfo
  {author} {\bibfnamefont {B.}~\bibnamefont {Hu}},\ }\bibfield  {title}
  {\enquote {\bibinfo {title} {Heat conduction in the nonlinear response
  regime: Scaling, boundary jumps, and negative differential thermal
  resistance},}\ }\href {\doibase 10.1103/PhysRevE.81.041131} {\bibfield
  {journal} {\bibinfo  {journal} {Phys. Rev. E}\ }\textbf {\bibinfo {volume}
  {81}},\ \bibinfo {pages} {041131} (\bibinfo {year} {2010})}\BibitemShut
  {NoStop}%
\bibitem [{\citenamefont {He}, \citenamefont {Buyukdagli},\ and\ \citenamefont
  {Hu}(2009)}]{PhysRevB.80.104302}%
  \BibitemOpen
  \bibfield  {author} {\bibinfo {author} {\bibfnamefont {D.}~\bibnamefont
  {He}}, \bibinfo {author} {\bibfnamefont {S.}~\bibnamefont {Buyukdagli}}, \
  and\ \bibinfo {author} {\bibfnamefont {B.}~\bibnamefont {Hu}},\ }\bibfield
  {title} {\enquote {\bibinfo {title} {Origin of negative differential thermal
  resistance in a chain of two weakly coupled nonlinear lattices},}\ }\href
  {\doibase 10.1103/PhysRevB.80.104302} {\bibfield  {journal} {\bibinfo
  {journal} {Phys. Rev. B}\ }\textbf {\bibinfo {volume} {80}},\ \bibinfo
  {pages} {104302} (\bibinfo {year} {2009})}\BibitemShut {NoStop}%
\bibitem [{\citenamefont {Hu}\ and\ \citenamefont
  {Chen}(2013)}]{PhysRevE.87.062104}%
  \BibitemOpen
  \bibfield  {author} {\bibinfo {author} {\bibfnamefont {J.}~\bibnamefont
  {Hu}}\ and\ \bibinfo {author} {\bibfnamefont {Y.~P.}\ \bibnamefont {Chen}},\
  }\bibfield  {title} {\enquote {\bibinfo {title} {Existence of negative
  differential thermal conductance in one-dimensional diffusive thermal
  transport},}\ }\href {\doibase 10.1103/PhysRevE.87.062104} {\bibfield
  {journal} {\bibinfo  {journal} {Phys. Rev. E}\ }\textbf {\bibinfo {volume}
  {87}},\ \bibinfo {pages} {062104} (\bibinfo {year} {2013})}\BibitemShut
  {NoStop}%
\bibitem [{\citenamefont {Li}\ \emph {et~al.}(2012)\citenamefont {Li},
  \citenamefont {Ren}, \citenamefont {Wang}, \citenamefont {Zhang},
  \citenamefont {H\"anggi},\ and\ \citenamefont {Li}}]{LiRMP84}%
  \BibitemOpen
  \bibfield  {author} {\bibinfo {author} {\bibfnamefont {N.}~\bibnamefont
  {Li}}, \bibinfo {author} {\bibfnamefont {J.}~\bibnamefont {Ren}}, \bibinfo
  {author} {\bibfnamefont {L.}~\bibnamefont {Wang}}, \bibinfo {author}
  {\bibfnamefont {G.}~\bibnamefont {Zhang}}, \bibinfo {author} {\bibfnamefont
  {P.}~\bibnamefont {H\"anggi}}, \ and\ \bibinfo {author} {\bibfnamefont
  {B.}~\bibnamefont {Li}},\ }\bibfield  {title} {\enquote {\bibinfo {title}
  {Colloquium: Phononics: Manipulating heat flow with electronic analogs and
  beyond},}\ }\href {\doibase 10.1103/RevModPhys.84.1045} {\bibfield  {journal}
  {\bibinfo  {journal} {Rev. Mod. Phys.}\ }\textbf {\bibinfo {volume} {84}},\
  \bibinfo {pages} {1045--1066} (\bibinfo {year} {2012})}\BibitemShut {NoStop}%
\bibitem [{\citenamefont {Fornieri}\ \emph {et~al.}(2016)\citenamefont
  {Fornieri}, \citenamefont {Timossi}, \citenamefont {Bosisio}, \citenamefont
  {Solinas},\ and\ \citenamefont {Giazotto}}]{FornieriPRB93}%
  \BibitemOpen
  \bibfield  {author} {\bibinfo {author} {\bibfnamefont {A.}~\bibnamefont
  {Fornieri}}, \bibinfo {author} {\bibfnamefont {G.}~\bibnamefont {Timossi}},
  \bibinfo {author} {\bibfnamefont {R.}~\bibnamefont {Bosisio}}, \bibinfo
  {author} {\bibfnamefont {P.}~\bibnamefont {Solinas}}, \ and\ \bibinfo
  {author} {\bibfnamefont {F.}~\bibnamefont {Giazotto}},\ }\bibfield  {title}
  {\enquote {\bibinfo {title} {Negative differential thermal conductance and
  heat amplification in superconducting hybrid devices},}\ }\href {\doibase
  10.1103/PhysRevB.93.134508} {\bibfield  {journal} {\bibinfo  {journal} {Phys.
  Rev. B}\ }\textbf {\bibinfo {volume} {93}},\ \bibinfo {pages} {134508}
  (\bibinfo {year} {2016})}\BibitemShut {NoStop}%
\bibitem [{\citenamefont {Zare}(2019)}]{Zare_2019}%
  \BibitemOpen
  \bibfield  {author} {\bibinfo {author} {\bibfnamefont {M.}~\bibnamefont
  {Zare}},\ }\bibfield  {title} {\enquote {\bibinfo {title} {Negative
  differential thermal conductance in a borophane normal
  metal{\textendash}superconductor junction},}\ }\href {\doibase
  10.1088/1361-6668/ab3caf} {\bibfield  {journal} {\bibinfo  {journal}
  {Supercond. Sci. Technol.}\ }\textbf {\bibinfo {volume} {32}},\ \bibinfo
  {pages} {115002} (\bibinfo {year} {2019})}\BibitemShut {NoStop}%
\bibitem [{\citenamefont {Sood}\ \emph {et~al.}(2018)\citenamefont {Sood},
  \citenamefont {Xiong}, \citenamefont {Chen}, \citenamefont {Wang},
  \citenamefont {Selli}, \citenamefont {Zhang}, \citenamefont {McClellan},
  \citenamefont {Sun}, \citenamefont {Donadio}, \citenamefont {Cui},
  \citenamefont {Pop},\ and\ \citenamefont {Goodson}}]{SoodNatComm2018}%
  \BibitemOpen
  \bibfield  {author} {\bibinfo {author} {\bibfnamefont {A.}~\bibnamefont
  {Sood}}, \bibinfo {author} {\bibfnamefont {F.}~\bibnamefont {Xiong}},
  \bibinfo {author} {\bibfnamefont {S.}~\bibnamefont {Chen}}, \bibinfo {author}
  {\bibfnamefont {H.}~\bibnamefont {Wang}}, \bibinfo {author} {\bibfnamefont
  {D.}~\bibnamefont {Selli}}, \bibinfo {author} {\bibfnamefont
  {J.}~\bibnamefont {Zhang}}, \bibinfo {author} {\bibfnamefont {C.~J.}\
  \bibnamefont {McClellan}}, \bibinfo {author} {\bibfnamefont {J.}~\bibnamefont
  {Sun}}, \bibinfo {author} {\bibfnamefont {D.}~\bibnamefont {Donadio}},
  \bibinfo {author} {\bibfnamefont {Y.}~\bibnamefont {Cui}}, \bibinfo {author}
  {\bibfnamefont {E.}~\bibnamefont {Pop}}, \ and\ \bibinfo {author}
  {\bibfnamefont {K.~E.}\ \bibnamefont {Goodson}},\ }\bibfield  {title}
  {\enquote {\bibinfo {title} {An electrochemical thermal transistor},}\ }\href
  {\doibase 10.1038/s41467-018-06760-7} {\bibfield  {journal} {\bibinfo
  {journal} {Nat. Commun.}\ }\textbf {\bibinfo {volume} {9}},\ \bibinfo {pages}
  {4510} (\bibinfo {year} {2018})}\BibitemShut {NoStop}%
\bibitem [{\citenamefont {Ben-Abdallah}\ and\ \citenamefont
  {Biehs}(2014)}]{Ben-Abdallah_PRL112}%
  \BibitemOpen
  \bibfield  {author} {\bibinfo {author} {\bibfnamefont {P.}~\bibnamefont
  {Ben-Abdallah}}\ and\ \bibinfo {author} {\bibfnamefont {S.-A.}\ \bibnamefont
  {Biehs}},\ }\bibfield  {title} {\enquote {\bibinfo {title} {Near-field
  thermal transistor},}\ }\href {\doibase 10.1103/PhysRevLett.112.044301}
  {\bibfield  {journal} {\bibinfo  {journal} {Phys. Rev. Lett.}\ }\textbf
  {\bibinfo {volume} {112}},\ \bibinfo {pages} {044301} (\bibinfo {year}
  {2014})}\BibitemShut {NoStop}%
\bibitem [{\citenamefont {Biehs}\ \emph {et~al.}(2021)\citenamefont {Biehs},
  \citenamefont {Messina}, \citenamefont {Venkataram}, \citenamefont
  {Rodriguez}, \citenamefont {Cuevas},\ and\ \citenamefont
  {Ben-Abdallah}}]{NearFieldRMP93}%
  \BibitemOpen
  \bibfield  {author} {\bibinfo {author} {\bibfnamefont {S.-A.}\ \bibnamefont
  {Biehs}}, \bibinfo {author} {\bibfnamefont {R.}~\bibnamefont {Messina}},
  \bibinfo {author} {\bibfnamefont {P.~S.}\ \bibnamefont {Venkataram}},
  \bibinfo {author} {\bibfnamefont {A.~W.}\ \bibnamefont {Rodriguez}}, \bibinfo
  {author} {\bibfnamefont {J.~C.}\ \bibnamefont {Cuevas}}, \ and\ \bibinfo
  {author} {\bibfnamefont {P.}~\bibnamefont {Ben-Abdallah}},\ }\bibfield
  {title} {\enquote {\bibinfo {title} {Near-field radiative heat transfer in
  many-body systems},}\ }\href {\doibase 10.1103/RevModPhys.93.025009}
  {\bibfield  {journal} {\bibinfo  {journal} {Rev. Mod. Phys.}\ }\textbf
  {\bibinfo {volume} {93}},\ \bibinfo {pages} {025009} (\bibinfo {year}
  {2021})}\BibitemShut {NoStop}%
\bibitem [{\citenamefont {Moncada-Villa}\ and\ \citenamefont
  {Cuevas}(2021)}]{Moncada-Villa_PRApplied15}%
  \BibitemOpen
  \bibfield  {author} {\bibinfo {author} {\bibfnamefont {E.}~\bibnamefont
  {Moncada-Villa}}\ and\ \bibinfo {author} {\bibfnamefont {J.~C.}\ \bibnamefont
  {Cuevas}},\ }\bibfield  {title} {\enquote {\bibinfo {title}
  {Normal-metal--superconductor near-field thermal diodes and transistors},}\
  }\href {\doibase 10.1103/PhysRevApplied.15.024036} {\bibfield  {journal}
  {\bibinfo  {journal} {Phys. Rev. Applied}\ }\textbf {\bibinfo {volume}
  {15}},\ \bibinfo {pages} {024036} (\bibinfo {year} {2021})}\BibitemShut
  {NoStop}%
\bibitem [{\citenamefont {Kats}\ \emph {et~al.}(2013)\citenamefont {Kats},
  \citenamefont {Blanchard}, \citenamefont {Zhang}, \citenamefont {Genevet},
  \citenamefont {Ko}, \citenamefont {Ramanathan},\ and\ \citenamefont
  {Capasso}}]{PhysRevX.3.041004}%
  \BibitemOpen
  \bibfield  {author} {\bibinfo {author} {\bibfnamefont {M.~A.}\ \bibnamefont
  {Kats}}, \bibinfo {author} {\bibfnamefont {R.}~\bibnamefont {Blanchard}},
  \bibinfo {author} {\bibfnamefont {S.}~\bibnamefont {Zhang}}, \bibinfo
  {author} {\bibfnamefont {P.}~\bibnamefont {Genevet}}, \bibinfo {author}
  {\bibfnamefont {C.}~\bibnamefont {Ko}}, \bibinfo {author} {\bibfnamefont
  {S.}~\bibnamefont {Ramanathan}}, \ and\ \bibinfo {author} {\bibfnamefont
  {F.}~\bibnamefont {Capasso}},\ }\bibfield  {title} {\enquote {\bibinfo
  {title} {Vanadium dioxide as a natural disordered metamaterial: Perfect
  thermal emission and large broadband negative differential thermal
  emittance},}\ }\href {\doibase 10.1103/PhysRevX.3.041004} {\bibfield
  {journal} {\bibinfo  {journal} {Phys. Rev. X}\ }\textbf {\bibinfo {volume}
  {3}},\ \bibinfo {pages} {041004} (\bibinfo {year} {2013})}\BibitemShut
  {NoStop}%
\bibitem [{\citenamefont {Ito}\ \emph {et~al.}(2014)\citenamefont {Ito},
  \citenamefont {Nishikawa}, \citenamefont {Iizuka},\ and\ \citenamefont
  {Toshiyoshi}}]{ItoAPL105}%
  \BibitemOpen
  \bibfield  {author} {\bibinfo {author} {\bibfnamefont {K.}~\bibnamefont
  {Ito}}, \bibinfo {author} {\bibfnamefont {K.}~\bibnamefont {Nishikawa}},
  \bibinfo {author} {\bibfnamefont {H.}~\bibnamefont {Iizuka}}, \ and\ \bibinfo
  {author} {\bibfnamefont {H.}~\bibnamefont {Toshiyoshi}},\ }\bibfield  {title}
  {\enquote {\bibinfo {title} {Experimental investigation of radiative thermal
  rectifier using vanadium dioxide},}\ }\href {\doibase 10.1063/1.4905132}
  {\bibfield  {journal} {\bibinfo  {journal} {Appl. Phys. Lett.}\ }\textbf
  {\bibinfo {volume} {105}},\ \bibinfo {pages} {253503} (\bibinfo {year}
  {2014})}\BibitemShut {NoStop}%
\bibitem [{\citenamefont {Musilov\'a}\ \emph {et~al.}(2019)\citenamefont
  {Musilov\'a}, \citenamefont {Kr\'al\'{\i}k}, \citenamefont
  {Fo\ifmmode~\check{r}\else \v{r}\fi{}t},\ and\ \citenamefont
  {Macek}}]{NbNPRB99}%
  \BibitemOpen
  \bibfield  {author} {\bibinfo {author} {\bibfnamefont {V.~c.~v.}\
  \bibnamefont {Musilov\'a}}, \bibinfo {author} {\bibfnamefont {T.~c.~v.}\
  \bibnamefont {Kr\'al\'{\i}k}}, \bibinfo {author} {\bibfnamefont {T.~c.~v.}\
  \bibnamefont {Fo\ifmmode~\check{r}\else \v{r}\fi{}t}}, \ and\ \bibinfo
  {author} {\bibfnamefont {M.}~\bibnamefont {Macek}},\ }\bibfield  {title}
  {\enquote {\bibinfo {title} {Strong suppression of near-field radiative heat
  transfer by superconductivity in nbn},}\ }\href {\doibase
  10.1103/PhysRevB.99.024511} {\bibfield  {journal} {\bibinfo  {journal} {Phys.
  Rev. B}\ }\textbf {\bibinfo {volume} {99}},\ \bibinfo {pages} {024511}
  (\bibinfo {year} {2019})}\BibitemShut {NoStop}%
\bibitem [{\citenamefont {Kikkert}(2013)}]{kikkert2013rf}%
  \BibitemOpen
  \bibfield  {author} {\bibinfo {author} {\bibfnamefont {C.~J.}\ \bibnamefont
  {Kikkert}},\ }\href@noop {} {\enquote {\bibinfo {title} {Rf electronics:
  design and simulation},}\ } (\bibinfo {year} {2013})\BibitemShut {NoStop}%
\bibitem [{\citenamefont {Meschke}, \citenamefont {Guichard},\ and\
  \citenamefont {Pekola}(2006)}]{MeschkePhotonic}%
  \BibitemOpen
  \bibfield  {author} {\bibinfo {author} {\bibfnamefont {M.}~\bibnamefont
  {Meschke}}, \bibinfo {author} {\bibfnamefont {W.}~\bibnamefont {Guichard}}, \
  and\ \bibinfo {author} {\bibfnamefont {J.~P.}\ \bibnamefont {Pekola}},\
  }\bibfield  {title} {\enquote {\bibinfo {title} {Single-mode heat conduction
  by photons},}\ }\href {\doibase 10.1038/nature05276} {\bibfield  {journal}
  {\bibinfo  {journal} {Nature}\ }\textbf {\bibinfo {volume} {444}},\ \bibinfo
  {pages} {187} (\bibinfo {year} {2006})}\BibitemShut {NoStop}%
\bibitem [{\citenamefont {Ronzani}\ \emph {et~al.}(2018)\citenamefont
  {Ronzani}, \citenamefont {Karimi}, \citenamefont {Senior}, \citenamefont
  {Chang}, \citenamefont {Peltonen}, \citenamefont {Chen},\ and\ \citenamefont
  {Pekola}}]{RonzaniPhotonic}%
  \BibitemOpen
  \bibfield  {author} {\bibinfo {author} {\bibfnamefont {A.}~\bibnamefont
  {Ronzani}}, \bibinfo {author} {\bibfnamefont {B.}~\bibnamefont {Karimi}},
  \bibinfo {author} {\bibfnamefont {J.}~\bibnamefont {Senior}}, \bibinfo
  {author} {\bibfnamefont {Y.-C.}\ \bibnamefont {Chang}}, \bibinfo {author}
  {\bibfnamefont {J.~T.}\ \bibnamefont {Peltonen}}, \bibinfo {author}
  {\bibfnamefont {C.}~\bibnamefont {Chen}}, \ and\ \bibinfo {author}
  {\bibfnamefont {J.~P.}\ \bibnamefont {Pekola}},\ }\bibfield  {title}
  {\enquote {\bibinfo {title} {Tunable photonic heat transport in a quantum
  heat valve},}\ }\href {\doibase 10.1038/s41567-018-0199-4} {\bibfield
  {journal} {\bibinfo  {journal} {Nat. Phys.}\ }\textbf {\bibinfo {volume}
  {14}},\ \bibinfo {pages} {991} (\bibinfo {year} {2018})}\BibitemShut
  {NoStop}%
\bibitem [{\citenamefont {Senior}\ \emph {et~al.}(2020)\citenamefont {Senior},
  \citenamefont {Gubaydullin}, \citenamefont {Karimi}, \citenamefont
  {Peltonen}, \citenamefont {Ankerhold},\ and\ \citenamefont
  {Pekola}}]{SeniorPhotonic}%
  \BibitemOpen
  \bibfield  {author} {\bibinfo {author} {\bibfnamefont {J.}~\bibnamefont
  {Senior}}, \bibinfo {author} {\bibfnamefont {A.}~\bibnamefont {Gubaydullin}},
  \bibinfo {author} {\bibfnamefont {B.}~\bibnamefont {Karimi}}, \bibinfo
  {author} {\bibfnamefont {J.~T.}\ \bibnamefont {Peltonen}}, \bibinfo {author}
  {\bibfnamefont {J.}~\bibnamefont {Ankerhold}}, \ and\ \bibinfo {author}
  {\bibfnamefont {J.~P.}\ \bibnamefont {Pekola}},\ }\bibfield  {title}
  {\enquote {\bibinfo {title} {Heat rectification via a superconducting
  artificial atom},}\ }\href {\doibase 10.1038/s42005-020-0307-5} {\bibfield
  {journal} {\bibinfo  {journal} {Commun. Phys.}\ }\textbf {\bibinfo {volume}
  {3}},\ \bibinfo {pages} {40} (\bibinfo {year} {2020})}\BibitemShut {NoStop}%
\bibitem [{\citenamefont {Maillet}\ \emph {et~al.}(2020)\citenamefont
  {Maillet}, \citenamefont {Subero}, \citenamefont {Peltonen}, \citenamefont
  {Golubev},\ and\ \citenamefont {Pekola}}]{MailletPhotonic}%
  \BibitemOpen
  \bibfield  {author} {\bibinfo {author} {\bibfnamefont {O.}~\bibnamefont
  {Maillet}}, \bibinfo {author} {\bibfnamefont {D.}~\bibnamefont {Subero}},
  \bibinfo {author} {\bibfnamefont {J.~T.}\ \bibnamefont {Peltonen}}, \bibinfo
  {author} {\bibfnamefont {D.~S.}\ \bibnamefont {Golubev}}, \ and\ \bibinfo
  {author} {\bibfnamefont {J.~P.}\ \bibnamefont {Pekola}},\ }\bibfield  {title}
  {\enquote {\bibinfo {title} {Electric field control of radiative heat
  transfer in a superconducting circuit},}\ }\href {\doibase
  10.1038/s41467-020-18163-8} {\bibfield  {journal} {\bibinfo  {journal} {Nat.
  Commun.}\ }\textbf {\bibinfo {volume} {11}},\ \bibinfo {pages} {4326}
  (\bibinfo {year} {2020})}\BibitemShut {NoStop}%
\bibitem [{\citenamefont {Thomas}, \citenamefont {Pekola},\ and\ \citenamefont
  {Golubev}(2019)}]{Thomas_PhotonicJJ_PRB100}%
  \BibitemOpen
  \bibfield  {author} {\bibinfo {author} {\bibfnamefont {G.}~\bibnamefont
  {Thomas}}, \bibinfo {author} {\bibfnamefont {J.~P.}\ \bibnamefont {Pekola}},
  \ and\ \bibinfo {author} {\bibfnamefont {D.~S.}\ \bibnamefont {Golubev}},\
  }\bibfield  {title} {\enquote {\bibinfo {title} {Photonic heat transport
  across a josephson junction},}\ }\href {\doibase 10.1103/PhysRevB.100.094508}
  {\bibfield  {journal} {\bibinfo  {journal} {Phys. Rev. B}\ }\textbf {\bibinfo
  {volume} {100}},\ \bibinfo {pages} {094508} (\bibinfo {year}
  {2019})}\BibitemShut {NoStop}%
\bibitem [{\citenamefont {Gubaydullin}\ \emph {et~al.}(2021)\citenamefont
  {Gubaydullin}, \citenamefont {Thomas}, \citenamefont {Golubev}, \citenamefont
  {Lvov}, \citenamefont {Peltonen},\ and\ \citenamefont
  {Pekola}}]{gubaydullin2021photonic}%
  \BibitemOpen
  \bibfield  {author} {\bibinfo {author} {\bibfnamefont {A.}~\bibnamefont
  {Gubaydullin}}, \bibinfo {author} {\bibfnamefont {G.}~\bibnamefont {Thomas}},
  \bibinfo {author} {\bibfnamefont {D.~S.}\ \bibnamefont {Golubev}}, \bibinfo
  {author} {\bibfnamefont {D.}~\bibnamefont {Lvov}}, \bibinfo {author}
  {\bibfnamefont {J.~T.}\ \bibnamefont {Peltonen}}, \ and\ \bibinfo {author}
  {\bibfnamefont {J.~P.}\ \bibnamefont {Pekola}},\ }\href@noop {} {\enquote
  {\bibinfo {title} {Photonic heat transport in three terminal superconducting
  circuit},}\ } (\bibinfo {year} {2021}),\ \Eprint
  {http://arxiv.org/abs/2112.09224} {arXiv:2112.09224 [cond-mat.mes-hall]}
  \BibitemShut {NoStop}%
\bibitem [{\citenamefont {Karimi}\ \emph {et~al.}(2017)\citenamefont {Karimi},
  \citenamefont {Pekola}, \citenamefont {Campisi},\ and\ \citenamefont
  {Fazio}}]{Karimi_2017}%
  \BibitemOpen
  \bibfield  {author} {\bibinfo {author} {\bibfnamefont {B.}~\bibnamefont
  {Karimi}}, \bibinfo {author} {\bibfnamefont {J.~P.}\ \bibnamefont {Pekola}},
  \bibinfo {author} {\bibfnamefont {M.}~\bibnamefont {Campisi}}, \ and\
  \bibinfo {author} {\bibfnamefont {R.}~\bibnamefont {Fazio}},\ }\bibfield
  {title} {\enquote {\bibinfo {title} {Coupled qubits as a quantum heat
  switch},}\ }\href {\doibase 10.1088/2058-9565/aa8330} {\bibfield  {journal}
  {\bibinfo  {journal} {Quantum Sci. Technol.}\ }\textbf {\bibinfo {volume}
  {2}},\ \bibinfo {pages} {044007} (\bibinfo {year} {2017})}\BibitemShut
  {NoStop}%
\bibitem [{\citenamefont {Nefzaoui}\ \emph {et~al.}(2014)\citenamefont
  {Nefzaoui}, \citenamefont {Joulain}, \citenamefont {Drevillon},\ and\
  \citenamefont {Ezzahri}}]{Nefzaoui2014}%
  \BibitemOpen
  \bibfield  {author} {\bibinfo {author} {\bibfnamefont {E.}~\bibnamefont
  {Nefzaoui}}, \bibinfo {author} {\bibfnamefont {K.}~\bibnamefont {Joulain}},
  \bibinfo {author} {\bibfnamefont {J.}~\bibnamefont {Drevillon}}, \ and\
  \bibinfo {author} {\bibfnamefont {Y.}~\bibnamefont {Ezzahri}},\ }\bibfield
  {title} {\enquote {\bibinfo {title} {Radiative thermal rectification using
  superconducting materials},}\ }\href {\doibase 10.1063/1.4868251} {\bibfield
  {journal} {\bibinfo  {journal} {Appl. Phys. Lett.}\ }\textbf {\bibinfo
  {volume} {104}},\ \bibinfo {pages} {103905} (\bibinfo {year}
  {2014})}\BibitemShut {NoStop}%
\bibitem [{\citenamefont {Bosisio}\ \emph {et~al.}(2016)\citenamefont
  {Bosisio}, \citenamefont {Solinas}, \citenamefont {Braggio},\ and\
  \citenamefont {Giazotto}}]{BosisioPRB93}%
  \BibitemOpen
  \bibfield  {author} {\bibinfo {author} {\bibfnamefont {R.}~\bibnamefont
  {Bosisio}}, \bibinfo {author} {\bibfnamefont {P.}~\bibnamefont {Solinas}},
  \bibinfo {author} {\bibfnamefont {A.}~\bibnamefont {Braggio}}, \ and\
  \bibinfo {author} {\bibfnamefont {F.}~\bibnamefont {Giazotto}},\ }\bibfield
  {title} {\enquote {\bibinfo {title} {Photonic heat conduction in
  josephson-coupled bardeen-cooper-schrieffer superconductors},}\ }\href
  {\doibase 10.1103/PhysRevB.93.144512} {\bibfield  {journal} {\bibinfo
  {journal} {Phys. Rev. B}\ }\textbf {\bibinfo {volume} {93}},\ \bibinfo
  {pages} {144512} (\bibinfo {year} {2016})}\BibitemShut {NoStop}%
\bibitem [{\citenamefont {Ordonez-Miranda}\ \emph {et~al.}(2017)\citenamefont
  {Ordonez-Miranda}, \citenamefont {Joulain}, \citenamefont {De~Sousa~Meneses},
  \citenamefont {Ezzahri},\ and\ \citenamefont
  {Drevillon}}]{Ordonez-Miranda2017}%
  \BibitemOpen
  \bibfield  {author} {\bibinfo {author} {\bibfnamefont {J.}~\bibnamefont
  {Ordonez-Miranda}}, \bibinfo {author} {\bibfnamefont {K.}~\bibnamefont
  {Joulain}}, \bibinfo {author} {\bibfnamefont {D.}~\bibnamefont
  {De~Sousa~Meneses}}, \bibinfo {author} {\bibfnamefont {Y.}~\bibnamefont
  {Ezzahri}}, \ and\ \bibinfo {author} {\bibfnamefont {J.}~\bibnamefont
  {Drevillon}},\ }\bibfield  {title} {\enquote {\bibinfo {title} {Photonic
  thermal diode based on superconductors},}\ }\href {\doibase
  10.1063/1.4991516} {\bibfield  {journal} {\bibinfo  {journal} {J. Appl.
  Phys.}\ }\textbf {\bibinfo {volume} {122}},\ \bibinfo {pages} {093105}
  (\bibinfo {year} {2017})}\BibitemShut {NoStop}%
\bibitem [{\citenamefont {Marchegiani}, \citenamefont {Braggio},\ and\
  \citenamefont {Giazotto}(2021)}]{MarchegianiAPLPhotonic}%
  \BibitemOpen
  \bibfield  {author} {\bibinfo {author} {\bibfnamefont {G.}~\bibnamefont
  {Marchegiani}}, \bibinfo {author} {\bibfnamefont {A.}~\bibnamefont
  {Braggio}}, \ and\ \bibinfo {author} {\bibfnamefont {F.}~\bibnamefont
  {Giazotto}},\ }\bibfield  {title} {\enquote {\bibinfo {title} {Highly
  efficient phase-tunable photonic thermal diode},}\ }\href {\doibase
  10.1063/5.0036485} {\bibfield  {journal} {\bibinfo  {journal} {Appl. Phys.
  Lett.}\ }\textbf {\bibinfo {volume} {118}},\ \bibinfo {pages} {022602}
  (\bibinfo {year} {2021})}\BibitemShut {NoStop}%
\bibitem [{Note1()}]{Note1}%
  \BibitemOpen
  \bibinfo {note} {At $T=1$K, the photon thermal wavelength is $\lambda
  =hc/(k_B T)=1.4$ cm.}\BibitemShut {Stop}%
\bibitem [{\citenamefont {Lifshitz}\ and\ \citenamefont
  {Pitaevskii}(2013)}]{lifshitz2013statistical}%
  \BibitemOpen
  \bibfield  {author} {\bibinfo {author} {\bibfnamefont {E.~M.}\ \bibnamefont
  {Lifshitz}}\ and\ \bibinfo {author} {\bibfnamefont {L.~P.}\ \bibnamefont
  {Pitaevskii}},\ }\href@noop {} {\emph {\bibinfo {title} {Statistical physics:
  theory of the condensed state}}},\ Vol.~\bibinfo {volume} {9}\ (\bibinfo
  {publisher} {Elsevier},\ \bibinfo {year} {2013})\BibitemShut {NoStop}%
\bibitem [{\citenamefont {Schmidt}, \citenamefont {Schoelkopf},\ and\
  \citenamefont {Cleland}(2004)}]{SchmidtPRL93}%
  \BibitemOpen
  \bibfield  {author} {\bibinfo {author} {\bibfnamefont {D.~R.}\ \bibnamefont
  {Schmidt}}, \bibinfo {author} {\bibfnamefont {R.~J.}\ \bibnamefont
  {Schoelkopf}}, \ and\ \bibinfo {author} {\bibfnamefont {A.~N.}\ \bibnamefont
  {Cleland}},\ }\bibfield  {title} {\enquote {\bibinfo {title} {Photon-mediated
  thermal relaxation of electrons in nanostructures},}\ }\href {\doibase
  10.1103/PhysRevLett.93.045901} {\bibfield  {journal} {\bibinfo  {journal}
  {Phys. Rev. Lett.}\ }\textbf {\bibinfo {volume} {93}},\ \bibinfo {pages}
  {045901} (\bibinfo {year} {2004})}\BibitemShut {NoStop}%
\bibitem [{\citenamefont {Ojanen}\ and\ \citenamefont
  {Jauho}(2008)}]{OjanenPRL100}%
  \BibitemOpen
  \bibfield  {author} {\bibinfo {author} {\bibfnamefont {T.}~\bibnamefont
  {Ojanen}}\ and\ \bibinfo {author} {\bibfnamefont {A.-P.}\ \bibnamefont
  {Jauho}},\ }\bibfield  {title} {\enquote {\bibinfo {title} {Mesoscopic photon
  heat transistor},}\ }\href {\doibase 10.1103/PhysRevLett.100.155902}
  {\bibfield  {journal} {\bibinfo  {journal} {Phys. Rev. Lett.}\ }\textbf
  {\bibinfo {volume} {100}},\ \bibinfo {pages} {155902} (\bibinfo {year}
  {2008})}\BibitemShut {NoStop}%
\bibitem [{\citenamefont {Pascal}, \citenamefont {Courtois},\ and\
  \citenamefont {Hekking}(2011)}]{PascalPRB83}%
  \BibitemOpen
  \bibfield  {author} {\bibinfo {author} {\bibfnamefont {L.~M.~A.}\
  \bibnamefont {Pascal}}, \bibinfo {author} {\bibfnamefont {H.}~\bibnamefont
  {Courtois}}, \ and\ \bibinfo {author} {\bibfnamefont {F.~W.~J.}\ \bibnamefont
  {Hekking}},\ }\bibfield  {title} {\enquote {\bibinfo {title} {Circuit
  approach to photonic heat transport},}\ }\href {\doibase
  10.1103/PhysRevB.83.125113} {\bibfield  {journal} {\bibinfo  {journal} {Phys.
  Rev. B}\ }\textbf {\bibinfo {volume} {83}},\ \bibinfo {pages} {125113}
  (\bibinfo {year} {2011})}\BibitemShut {NoStop}%
\bibitem [{\citenamefont {Harrison}(2013)}]{Harrison2013}%
  \BibitemOpen
  \bibfield  {author} {\bibinfo {author} {\bibfnamefont {M.}~\bibnamefont
  {Harrison}},\ }\bibfield  {title} {\enquote {\bibinfo {title} {Physical
  collisions and the maximum power theorem: an analogy between mechanical and
  electrical situations},}\ }\href {\doibase 10.1088/0031-9120/48/2/207}
  {\bibfield  {journal} {\bibinfo  {journal} {Phys. Educ.}\ }\textbf {\bibinfo
  {volume} {48}},\ \bibinfo {pages} {207--211} (\bibinfo {year}
  {2013})}\BibitemShut {NoStop}%
\bibitem [{\citenamefont {Atkin}(2013)}]{Atkin2013}%
  \BibitemOpen
  \bibfield  {author} {\bibinfo {author} {\bibfnamefont {K.}~\bibnamefont
  {Atkin}},\ }\bibfield  {title} {\enquote {\bibinfo {title} {Energy transfer
  and a recurring mathematical function},}\ }\href {\doibase
  10.1088/0031-9120/48/5/616} {\bibfield  {journal} {\bibinfo  {journal} {Phys.
  Educ.}\ }\textbf {\bibinfo {volume} {48}},\ \bibinfo {pages} {616--620}
  (\bibinfo {year} {2013})}\BibitemShut {NoStop}%
\bibitem [{Note2()}]{Note2}%
  \BibitemOpen
  \bibinfo {note} {For simplicity, we neglect the superconducting proximity
  effect between the two elements~\cite {Pannetier2000}.}\BibitemShut {Stop}%
\bibitem [{\citenamefont {Steele}\ and\ \citenamefont
  {Hein}(1953)}]{TitaniumPhysRev92}%
  \BibitemOpen
  \bibfield  {author} {\bibinfo {author} {\bibfnamefont {M.~C.}\ \bibnamefont
  {Steele}}\ and\ \bibinfo {author} {\bibfnamefont {R.~A.}\ \bibnamefont
  {Hein}},\ }\bibfield  {title} {\enquote {\bibinfo {title} {Superconductivity
  of titanium},}\ }\href {\doibase 10.1103/PhysRev.92.243} {\bibfield
  {journal} {\bibinfo  {journal} {Phys. Rev.}\ }\textbf {\bibinfo {volume}
  {92}},\ \bibinfo {pages} {243--247} (\bibinfo {year} {1953})}\BibitemShut
  {NoStop}%
\bibitem [{\citenamefont {Thiemann}, \citenamefont {Dressel},\ and\
  \citenamefont {Scheffler}(2018)}]{DresselTitanium}%
  \BibitemOpen
  \bibfield  {author} {\bibinfo {author} {\bibfnamefont {M.}~\bibnamefont
  {Thiemann}}, \bibinfo {author} {\bibfnamefont {M.}~\bibnamefont {Dressel}}, \
  and\ \bibinfo {author} {\bibfnamefont {M.}~\bibnamefont {Scheffler}},\
  }\bibfield  {title} {\enquote {\bibinfo {title} {Complete electrodynamics of
  a bcs superconductor with $\ensuremath{\mu}\mathrm{eV}$ energy scales:
  Microwave spectroscopy on titanium at mk temperatures},}\ }\href {\doibase
  10.1103/PhysRevB.97.214516} {\bibfield  {journal} {\bibinfo  {journal} {Phys.
  Rev. B}\ }\textbf {\bibinfo {volume} {97}},\ \bibinfo {pages} {214516}
  (\bibinfo {year} {2018})}\BibitemShut {NoStop}%
\bibitem [{Note3()}]{Note3}%
  \BibitemOpen
  \bibinfo {note} {The bulk critical temperature of Al is $T_c^{\protect \rm
  Al}=1.2$ K and larger for thin films (up to $3K$ for 3 nm thick
  films).}\BibitemShut {Stop}%
\bibitem [{Note4()}]{Note4}%
  \BibitemOpen
  \bibinfo {note} {The dissipation in the wire is exponentially suppressed at
  low temperatures due to the presence of the gap. The kinetic inductance
  contribution can be minimized by geometry design. At the same time, direct
  conduction of heat between the electrodes and the wire is suppressed by
  Andreev mirroring.~\cite {MarchegianiAPLPhotonic}}\BibitemShut {NoStop}%
\bibitem [{\citenamefont {Mattis}\ and\ \citenamefont
  {Bardeen}(1958)}]{MattisBardeenPhysRev111}%
  \BibitemOpen
  \bibfield  {author} {\bibinfo {author} {\bibfnamefont {D.~C.}\ \bibnamefont
  {Mattis}}\ and\ \bibinfo {author} {\bibfnamefont {J.}~\bibnamefont
  {Bardeen}},\ }\bibfield  {title} {\enquote {\bibinfo {title} {Theory of the
  anomalous skin effect in normal and superconducting metals},}\ }\href
  {\doibase 10.1103/PhysRev.111.412} {\bibfield  {journal} {\bibinfo  {journal}
  {Phys. Rev.}\ }\textbf {\bibinfo {volume} {111}},\ \bibinfo {pages}
  {412--417} (\bibinfo {year} {1958})}\BibitemShut {NoStop}%
\bibitem [{Note5()}]{Note5}%
  \BibitemOpen
  \bibinfo {note} {In Eq.~\protect \textup {\hbox {\mathsurround \z@ \protect
  \normalfont (\ignorespaces \ref {eq:MBimag}\unskip \@@italiccorr )}},
  $\protect \sqrt {E^2-\Delta ^2}$ is purely imaginary and should be intended
  as $\protect \sqrt {\Delta ^2-E^2}$ due to the presence of the absolute
  value. The lower limit of integration is the maximum between $-\Delta $ and
  $\Delta -\omega $.}\BibitemShut {Stop}%
\bibitem [{Note6()}]{Note6}%
  \BibitemOpen
  \bibinfo {note} {This approximate expression gives an accurate description
  for the temperature dependence of the gap, with an error smaller than 3\% for
  every temperature.}\BibitemShut {Stop}%
\bibitem [{Note7()}]{Note7}%
  \BibitemOpen
  \bibinfo {note} {The cutoff at $2\Delta _0$ is related to the fact that we
  are mainly interested in temperatures $T_S\leq T_c$.}\BibitemShut {Stop}%
\bibitem [{\citenamefont {Timofeev}\ \emph {et~al.}(2009)\citenamefont
  {Timofeev}, \citenamefont {Garc\'{\i}a}, \citenamefont {Kopnin},
  \citenamefont {Savin}, \citenamefont {Meschke}, \citenamefont {Giazotto},\
  and\ \citenamefont {Pekola}}]{TimofeevPRL102}%
  \BibitemOpen
  \bibfield  {author} {\bibinfo {author} {\bibfnamefont {A.~V.}\ \bibnamefont
  {Timofeev}}, \bibinfo {author} {\bibfnamefont {C.~P.}\ \bibnamefont
  {Garc\'{\i}a}}, \bibinfo {author} {\bibfnamefont {N.~B.}\ \bibnamefont
  {Kopnin}}, \bibinfo {author} {\bibfnamefont {A.~M.}\ \bibnamefont {Savin}},
  \bibinfo {author} {\bibfnamefont {M.}~\bibnamefont {Meschke}}, \bibinfo
  {author} {\bibfnamefont {F.}~\bibnamefont {Giazotto}}, \ and\ \bibinfo
  {author} {\bibfnamefont {J.~P.}\ \bibnamefont {Pekola}},\ }\bibfield  {title}
  {\enquote {\bibinfo {title} {Recombination-limited energy relaxation in a
  bardeen-cooper-schrieffer superconductor},}\ }\href {\doibase
  10.1103/PhysRevLett.102.017003} {\bibfield  {journal} {\bibinfo  {journal}
  {Phys. Rev. Lett.}\ }\textbf {\bibinfo {volume} {102}},\ \bibinfo {pages}
  {017003} (\bibinfo {year} {2009})}\BibitemShut {NoStop}%
\bibitem [{\citenamefont {Maisi}\ \emph {et~al.}(2013)\citenamefont {Maisi},
  \citenamefont {Lotkhov}, \citenamefont {Kemppinen}, \citenamefont {Heimes},
  \citenamefont {Muhonen},\ and\ \citenamefont {Pekola}}]{MaisiPRL111}%
  \BibitemOpen
  \bibfield  {author} {\bibinfo {author} {\bibfnamefont {V.~F.}\ \bibnamefont
  {Maisi}}, \bibinfo {author} {\bibfnamefont {S.~V.}\ \bibnamefont {Lotkhov}},
  \bibinfo {author} {\bibfnamefont {A.}~\bibnamefont {Kemppinen}}, \bibinfo
  {author} {\bibfnamefont {A.}~\bibnamefont {Heimes}}, \bibinfo {author}
  {\bibfnamefont {J.~T.}\ \bibnamefont {Muhonen}}, \ and\ \bibinfo {author}
  {\bibfnamefont {J.~P.}\ \bibnamefont {Pekola}},\ }\bibfield  {title}
  {\enquote {\bibinfo {title} {Excitation of single quasiparticles in a small
  superconducting al island connected to normal-metal leads by tunnel
  junctions},}\ }\href {\doibase 10.1103/PhysRevLett.111.147001} {\bibfield
  {journal} {\bibinfo  {journal} {Phys. Rev. Lett.}\ }\textbf {\bibinfo
  {volume} {111}},\ \bibinfo {pages} {147001} (\bibinfo {year}
  {2013})}\BibitemShut {NoStop}%
\bibitem [{\citenamefont {Viisanen}\ and\ \citenamefont
  {Pekola}(2018)}]{ViisanenPRB97}%
  \BibitemOpen
  \bibfield  {author} {\bibinfo {author} {\bibfnamefont {K.~L.}\ \bibnamefont
  {Viisanen}}\ and\ \bibinfo {author} {\bibfnamefont {J.~P.}\ \bibnamefont
  {Pekola}},\ }\bibfield  {title} {\enquote {\bibinfo {title} {Anomalous
  electronic heat capacity of copper nanowires at sub-kelvin temperatures},}\
  }\href {\doibase 10.1103/PhysRevB.97.115422} {\bibfield  {journal} {\bibinfo
  {journal} {Phys. Rev. B}\ }\textbf {\bibinfo {volume} {97}},\ \bibinfo
  {pages} {115422} (\bibinfo {year} {2018})}\BibitemShut {NoStop}%
\bibitem [{\citenamefont {De~Simoni}\ \emph {et~al.}(2018)\citenamefont
  {De~Simoni}, \citenamefont {Paolucci}, \citenamefont {Solinas}, \citenamefont
  {Strambini},\ and\ \citenamefont {Giazotto}}]{DeSimoniGate}%
  \BibitemOpen
  \bibfield  {author} {\bibinfo {author} {\bibfnamefont {G.}~\bibnamefont
  {De~Simoni}}, \bibinfo {author} {\bibfnamefont {F.}~\bibnamefont {Paolucci}},
  \bibinfo {author} {\bibfnamefont {P.}~\bibnamefont {Solinas}}, \bibinfo
  {author} {\bibfnamefont {E.}~\bibnamefont {Strambini}}, \ and\ \bibinfo
  {author} {\bibfnamefont {F.}~\bibnamefont {Giazotto}},\ }\bibfield  {title}
  {\enquote {\bibinfo {title} {Metallic supercurrent field-effect
  transistor},}\ }\href {\doibase 10.1038/s41565-018-0190-3} {\bibfield
  {journal} {\bibinfo  {journal} {Nat. Nanotechnol.}\ }\textbf {\bibinfo
  {volume} {13}},\ \bibinfo {pages} {802} (\bibinfo {year} {2018})}\BibitemShut
  {NoStop}%
\bibitem [{\citenamefont {Pannetier}\ and\ \citenamefont
  {Courtois}(2000)}]{Pannetier2000}%
  \BibitemOpen
  \bibfield  {author} {\bibinfo {author} {\bibfnamefont {B.}~\bibnamefont
  {Pannetier}}\ and\ \bibinfo {author} {\bibfnamefont {H.}~\bibnamefont
  {Courtois}},\ }\bibfield  {title} {\enquote {\bibinfo {title} {Andreev
  reflection and proximity effect},}\ }\href {\doibase 10.1023/A:1004635226825}
  {\bibfield  {journal} {\bibinfo  {journal} {J. Low Temp. Phys.}\ }\textbf
  {\bibinfo {volume} {118}},\ \bibinfo {pages} {599--615} (\bibinfo {year}
  {2000})}\BibitemShut {NoStop}%
\end{thebibliography}
\end{document}